\documentclass[%
  tightenlines,
superscriptaddress,
 amsmath,amssymb,
 prl,
12pt]{revtex4-1}

\usepackage{graphicx}
\usepackage{dcolumn}
\usepackage{bm}
\usepackage[compact]{titlesec}
 \titlespacing{\section}{0pt}{2ex}{1ex}
    \titlespacing{\subsection}{0pt}{1ex}{0ex}
    \titlespacing{\subsubsection}{0pt}{0.5ex}{0ex}
\usepackage{mathptmx}
\usepackage[text={16cm,24cm}]{geometry}
\usepackage{color}

\newcommand{\msun}{{\rm M}_\odot}

\newcommand\lsim{\mathrel{\rlap{\lower4pt\hbox{\hskip1pt$\sim$}}
        \raise1pt\hbox{$<$}}}
\newcommand\gsim{\mathrel{\rlap{\lower4pt\hbox{\hskip1pt$\sim$}}
        \raise1pt\hbox{$>$}}}

\begin{document}

\noindent
\begin{center}
Primary Thematic Science Area: Multi-Messenger Astronomy and Astrophysics\\
Secondary Areas: Cosmology and Fundamental Physics, Galaxy Evolution,\\
Formation and Evolution of Compact Objects
\end{center}

\title{Multimessenger science opportunities with mHz gravitational waves}

\author{John Baker}
\affiliation{ NASA Goddard Space Flight Center, Greenbelt, MD 20771, USA}
\affiliation{ email: john.g.baker@nasa.gov}
\author{Zolt\'an Haiman}%
\affiliation{ Columbia University}
\author{Elena Maria Rossi}%
\affiliation{ Leiden University}
\author{\\Edo Berger}
\affiliation{ Harvard University}
\author{Niel Brandt}
\affiliation{ The Pennsylvania State University}
\author{Elm\'e Breedt}
\affiliation{ University of Cambridge}
\author{Katelyn Breivik}
\affiliation{Canadian Institute for Theoretical Astrophysics, University of Toronto}
\author{Maria Charisi}
\affiliation{ California Institute of Technology}
\author{Andrea Derdzinski}
\affiliation{ Columbia University}
\author{Daniel J. D'Orazio}
\affiliation{ Harvard University}
\author{Saavik Ford}
\affiliation{ City University of New York}
\affiliation{ American Museum of Natural History}
\author{Jenny E. Greene}
\affiliation{ Princeton University}
\author{J. Colin Hill}
\affiliation{ Institute for Advanced Study}
\affiliation{ Center for Computational Astrophysics, Flatiron Institute}
\author{Kelly Holley-Bockelmann}
\affiliation{ Vanderbilt University}
\author{Joey Shapiro Key}
\affiliation{ University of Washington, Bothell}
\author{Bence Kocsis}
\affiliation{ E\"otv\"os University}
\author{Thomas Kupfer}
\affiliation{ Kavli Institute for Theoretical Physics}
\author{Shane Larson}
\affiliation{ Northwestern University}
\author{Piero Madau}
\affiliation{ University of California, Santa Cruz}
\author{Thomas Marsh}
\affiliation{ University of Warwick}
\author{Barry McKernan}
\affiliation{ City University of New York}
\affiliation{ American Museum of Natural History}
\author{Sean T. McWilliams}
\affiliation{ West Virginia University}
\author{Priyamvada Natarajan}
\affiliation{ Yale University}
\author{Samaya Nissanke}
\affiliation{ GRAPPA, University of Amsterdam}
\author{Scott Noble}
\affiliation{ University of Tulsa}
\affiliation{ NASA Goddard Space Flight Center, Greenbelt, MD 20771, USA}
\author{E. Sterl Phinney}
\affiliation{ California Institute of Technology}
\author{Gavin Ramsay}
\affiliation{ Armagh Observatory}
\author{Jeremy Schnittman}
\affiliation{ NASA Goddard Space Flight Center, Greenbelt, MD 20771, USA}
\author{Alberto Sesana}
\affiliation{ University of Birmingham}
\affiliation{ Universit\`a di Milano Bicocca}
\author{David Shoemaker}
\affiliation{ LIGO, Massachusetts Institute of Technology}
\author{Nicholas Stone}
\affiliation{ Columbia University}
\author{Silvia Toonen}
\affiliation{ University of Amsterdam}
\affiliation{ University of Birmingham}
\author{Benny Trakhtenbrot}
\affiliation{ Tel Aviv University}
\author{Alexey Vikhlinin}
\affiliation{ Harvard University}
\author{Marta Volonteri}
\affiliation{ Institut d`Astrophysique de Paris}


\thispagestyle{empty}
\date{\today}



\maketitle


  \noindent{\it LISA will open the mHz band of gravitational waves (GWs) to the
  astronomy community.  The strong gravity which powers the variety of
  GW sources in this band is also crucial in a number of important
  astrophysical processes at the current frontiers of astronomy.
  These range from the beginning of structure formation in the early
  universe, through the origin and cosmic evolution of
  massive black holes in concert with their galactic environments, to
  the evolution of stellar remnant binaries in the
  Milky Way and in nearby galaxies.  These processes and their
  associated populations also drive current and future observations across the electromagnetic (EM)
  spectrum.  We review opportunities for science breakthroughs,
  involving either direct coincident EM+GW observations, or indirect
  multimessenger studies. We argue that for the US community to fully capitalize on the opportunities from the LISA mission, the US efforts should be accompanied by a coordinated and sustained 
  program of multi-disciplinary science investment, following the GW data through to its impact on broad
  areas of astrophysics.  Support for LISA-related multimessenger observers and
  theorists should be sized appropriately for a flagship observatory
  and may be coordinated through a dedicated mHz GW research center.}
  
\setcounter{page}{1}
\section{\label{sec:frontiers} Multimessenger Frontiers}
We begin by enumerating key science questions requiring {\em both} EM and GW measurements.  This serves as a mere snapshot of
current knowledge; since
the field is in its infancy, new discoveries will drive
multi-messenger science over the next decade, and we must be responsive.

{\bf Is General Relativity (GR) the Correct Theory of Gravity?}
While GR has passed a multitude of tests~\cite{will06,2018arXiv181100364T}, viable alternative theories of gravity remain, including some that
may account for apparent cosmic acceleration without invoking dark
energy.  Together with EM observations, LISA can test these theories
by determining the luminosity distance $d_L(z)$ of a high-SNR massive ($M\approx 10^6\msun$) black
hole (MBH) binary~\cite{Schutz86} with a few \% accuracy out to redshift $z\approx 3$~\cite{Holz_2005}.  A degeneracy between $M$ and $(1+z)$ in the GW waveform
precludes measuring the redshift of such a ``standard siren'', but if an
EM counterpart of the source, or its host galaxy, is identified, its
redshift can be measured.  GW distances are based on the
propagation of gravity, rather than light. A comparison of
standard sirens and standard EM candles at similar redshifts can therefore
probe alternative theories of gravity (e.g. with extra dimensions)
for which the GW and EM luminosity distances would disagree \cite{DefMen07}.
Similarly, in massive gravity theories, the speed of GWs 
differs from the speed of light, causing differences in the arrival
times from cosmological sources. Along with LIGO binary neutron star observations, which so far constrain this speed difference to $10^{-15}$,  mHz
sirens seen by LISA will provide new tests for theories with
frequency-dependent delays~\cite{deRham2018}, and potentially stronger tests
with improved systematics if the EM emission time can be well-understood theoretically~\cite{KHM2008} or if the EM chirp from merging binaries embedded in circumbinary gas~\citep{Haiman2017,Tang+2017,Bowen+2018} is phased with the GW chirp.
The GW distance-redshift relation is highly complementary to
that from supernovae (SNe):
(1) it side-steps the calibration via a
distance ladder, (2) has entirely different systematics, and (3)
will extend to redshifts inaccessible to SN
observations, easily probing dark energy models that
affect $d_L$ at $z>3$~\citep{tamanini16}. 
GW+EM standard sirens would also naturally resolve the current tensions about local {\it vs.} high-$z$ measurements of the Hubble constant ${\rm H_0}$~\citep{Feeney+2019}.

{\bf Did the Early Universe have an Inflationary Stage?}  
On large scales, our universe is dominated by dark energy and dark matter, with
primordial perturbations with a nearly 
scale-invariant power spectrum, consistent with
inflation~\cite{Weinberg2013,Planck2018-legacy,Planck2018-cosmology}.
Standard inflationary models generically predict a stochastic GW background (GWB),
whose power spectrum declines towards high frequency in the simplest, single-field, slow-roll models, though other models allow a range of spectral slopes.   The GWB can be probed at very low frequencies ($10^{-16}-10^{-18}$Hz) by cosmic microwave background
polarization~\citep{Bicep2/Planck,Calabrese-AdvACT,Arnold-Simons,Benson-SPT3G,SimonsObservatory}, and by gravitational waves at $\sim$nHz (by 
Pulsar timing arrays; PTAs~\cite{Verbiest+2016}),  $\sim$mHz (by LISA~\cite{Lasky+2016}) and  $\sim$100Hz (ground-based GW detectors) frequencies. The potential discovery of a GWB and a measurement (or constraint) of the slope of its spectrum over the enormous frequency range provided by the joint GW+EM observations would provide smoking gun evidence for inflation, as well as a powerful probe of non-minimal inflationary models.

  {\bf How do Supermassive Black Holes (SMBHs) and Galaxies
    Co-Evolve?}  Nuclear SMBHs correlate with many fundamental properties of their
  host galaxies, indicating that SMBHs and galaxies co-evolve over
  cosmic time~\cite{Yang+2019}. However, the nature of this co-evolution and the physics
  responsible for it is not yet understood~\cite{Kormendy2013}.  If EM
  observations can identify unique host galaxies of LISA SMBH
  binaries, then this will directly provide the relation between
  (merging) SMBHs and their host galaxies as a function of redshift,
  luminosity and other properties.  GW data will yield
  precise and reliable estimates of the masses, spins, and orbits of the SMBHs, which are not attainable from
  EM observations alone.

  {\bf How Did the Earliest SMBHs Form and Grow?}  One of the puzzling
  discoveries of the past two decades is the existence of billion-solar
  mass SMBHs in the first Gyr of the universe.  The early emergence
  of these SMBHs requires a rapid assembly of the first $\sim 10^6\msun$ SMBHs by $z>10$.  LISA will probe
  the assembly of these early SMBHs via mergers out to $z\sim 20$,
  while EM instruments pushing the sensitivity limits, such as Lynx, Athena, and JWST will reveal their growth by directly detecting the light
  produced in the X-ray and optical bands during their growth by
  accretion. These two observations together bracket the two main channels in
  which SMBHs can grow in mass over time: mergers or accretion.

  {\bf Do IMBHs Exist? If So, How Did They Form?}  Astronomers have
  found examples of stellar-mass BHs in the range $10^{0-2}\msun$, and
  SMBHs with $10^{6-10}\msun$.  It is currently unknown whether
  intermediate-mass BHs (IMBHs) exist in-between these two
  populations.  LISA and deep EM observations~\cite{Mezcua2017} will cover this range
  and together will be able to map out the demography and growth via
  mergers and accretion of the population, if any, of IMBHs.  While LISA
  could establish the mere existence of IMBHs (via an IMBH-IMBH merger
  or a stellar mass black hole spiraling into a larger IMBH), a white dwarf inspiralling into an IMBH
  would be tidally stripped and produce a unique bright EM
  counterpart~\citep{Sesana+2008b,Zalamea+2010}.  Host galaxy localization could
 help distinguish between different IMBH formation scenarios (such as
  runaway in globular clusters, formation in dwarf galaxies, or via
  copious accretion).

  {\bf How Do Accretion Disks Fuel AGN?}  AGN disks
  contribute
  to galaxy formation via various forms of feedback \citep{Somerville08}. While EM observations probe the photospheres of AGN disks, even basic disk properties, such as density, temperature, geometry, accretion rate or lifetime, remain poorly understood~\citep{Martini01}. GW signatures of mergers involving SMBHs and/or stellar-mass BHs in an AGN disk can complement EM data and provide novel information on disk properties in several ways. 
  Coupled with LISA observations, we will, for the first time, study bright EM emission from SMBHs whose masses, spins, and orbital parameters are \emph{precisely known}.
  LISA may also be able to directly measure the effect of gas drag on the GW waveform of a stellar-mass black hole (sBH) merging with the SMBH~\citep{LISA-IMRI-gas}.  The magnitude and frequency-dependence of the deviation from vacuum waveforms will probe disk properties and can be disentangled from uncertainties in the binary parameters~\citep{Kocsis_2011,Yunes_2011,Yunes_2011b,LISA-IMRI-gas}. 
  Stellar-mass BH binaries detectable by both LIGO and LISA~\citep{Sesana2016} may be captured~\citep{Bartos17} or produced~\citep{McKernan+2014,Stone17,Bellovary+2016} within AGN disks, as a consequence of nuclear star-formation. These possibilites may be explored with  multimessenger cross-catalogs~\citep{Inayoshi+2017,Meiron+2017,RandallXianyu2018}.

  {\bf How does Circumbinary Gas Interact with Binary Evolution?}
  The coupled behavior of binaries interacting with a circumbinary disk
  is still poorly understood despite decades of theoretical work, and
  despite its obvious relevance for (proto-)planetary disks and
  stellar binaries, in addition to SMBH binaries.  For example,
  even the sign of the disk torques (i.e. whether the circumbinary gas
  promotes inspiral or causes an outspiral) is being debated, both for
  equal-mass binaries \cite{Tang+2017,Munoz+2018,Moody+2019} and in the highly
  unequal-mass regime \cite{Baruteau+2014}.  
  LISA will measure the imprints of these interactions on the spin orientations and eccentricities of both binary SMBH and sBH-SMBH mergers.
  LISA observations may be complemented by EM observations of broad emission lines in the both X-ray and optical bands.  The presence of a binary will strongly perturb the nearby gas disk and will change the emission line kinematics and generate unusual periodic variability.  These effects should be especially large in the X-rays, probing gas closest to the SMBHs~\cite{d'Ascoli+2018}.  Broad X-ray lines, like Fe K$_{\alpha}$, can display strongly disturbed spectral shapes, and/or periodic Doppler modulations on the binary's orbital timescale of O(hr)~\citep{McK13,Sesana+2012,McK15}. 



{\bf What Are The Properties of Nuclear Stellar Bulges?}
  The demography and properties of dense star clusters in the centers of galaxies remain poorly understood.  EM observations of the tidal
  distruption of stars, together with GW observations of extreme mass ratio inspirals, will provide complementary insights about these systems\citep{EracleousWP}.  Stars passing within
  $\sim 10$ Schwarzschild radii of a SMBH are torn apart by its
  tide, and the accretion of the resulting stellar debris produces a luminous transient \citep{Rees88} visible 
  from radio to soft gamma-rays~\citep{Zauderer+11,vanVelzen+16,Gezari+12,Gezari+06,Komossa&Bade99,Bloom+11}.  While TDEs are rare around single SMBHs, their rates
  may be strongly enhanced in the scrambled environments of SMBH binaries, potentially generating a burst of TDEs associated with a LISA SMBH binary~\cite{Chen+2009}. 
  
  {\bf What is the Formation Mechanism and Fate of Stellar Binaries?}  
  Our understanding of stellar evolution and the stellar initial mass function imply that $>$95\% of stars in the Milky Way end up as white dwarfs (WDs). Between $5-10\%$ of WDs are in double WD (DWD) binaries \citep{maxted99,toonen17,maoz18}, implying the presence of $>10^8$ DWD systems~\citep{nelemans2004}. EM observations have revealed $\sim 150$ DWDs~\cite{napi03,brownw10}. The tightest of these (period $<1$hr) have experienced at least two phases of mass transfer and thus provide stringent tests for theories of binary evolution. Moreover, they constitute a formation channel for type Ia SN (SN Ia) \citep{webbink84,iben84}. Tight DWD binaries are abundant mHz GW sources, though not present in the ground-based GW waveband. Gaia and LSST should increase the number of guaranteed LISA sources by an order of magnitude from $\approx$10 currently known. As many as {\em tens of thousands} will be individually resolved by LISA \citep{korol17,breivik18,kupfer18}. This abundance of compact DWDs will enable unprecedented studies of their formation channels and demography 
  \citep[][]{toonen12}. Likewise, we can put on firm statistics the physical conditions that determine their fate. While candidate SN Ia progenitors from the double degenerate channel are hard to detect optically because of their small stellar size and separation~\footnote{The most promising candidate is Heinze 2-428 \citep{sg15} but see \citep{gb16}.} they are extremely bright in GWs. Indeed, GWs may be the only way to open an observational window onto this specific population \citep{Nissanke+2012,korolk18,rebassa18}. Besides DWDs, more exotic compact object binaries, such as white dwarf-neutron star \citep{tauris18} and double neutron star systems are expected at 2 to 3 orders of magnitude lower abundance \citep{nelemans2004}.  These systems may shed light on core-collapse and electron-capture SNe and related remnant recoils.  Optimal science return will require maximal effort to identify EM counterparts.  The {\it joint} EM-GW data analysis for individual binaries will allow us to nail down binary properties, such as individual masses, separation and the physical conditions governing the interaction between star members. 


  {\bf What is the Physics of Mass Transfer in DWDs?}
  GW-driven evolution may bring DWDs into contact. If the DWD survives the onset of mass-transfer to accrete stably, it is known as an AM\,CVn type
  binary.   Their number and properties provide an observational anchor for our understanding of WD mass transfer physics. 
  EM observations -- from optical to UV and X-ray -- can provide us with the stellar masses, orbital period, sky position, accretion luminosity, spectra, whereas GW measurements can provide sky position, orbital period (and period derivative), distance and the combined mass of the system (chirp mass). Armed with these measurements we can tackle the following questions on at least several tens of objects:
{\em ({\it i}) When does stable mass transfer occur in DWDs?}   
The largest uncertainty about the fate of DWDs (merger {\it vs.} stable mass transfer), and hence of SN Ia rates, is the onset of mass transfer when the larger WD starts to spill onto its companion. The observed number of systems is orders of magnitude below predictions~\cite{carter13,ramsay18}.
LISA DWDs will directly reveal how many systems avoid merger, and the properties of the surviving AM\,CVn binaries.
LISA will discover DWDs at their shortest periods, probing, with EM follow-up, the physics of interactions such as tidal braking, dynamical tides, resonance locking, and tidal heating~\citep{Marsh+2004M,Marsh2011,Fuller+2013,Fuller+2014}.
%
{\em ({\it ii}) How is angular momentum transported in accreting DWDs?}  
The properties derived from combined EM+GW data can disentangle the contribution from GWs and mass transfer to the period evolution ($\dot{P}$) and enable the first study of angular momentum transport in accreting DWDs in a statistically significant sample \cite{Breivik+2018}.
{\em ({\it iii}) What is the radiative efficiency of accreting WD systems?}
Angular momentum transport is intimately linked to the accretion rate
of a system.
Combined with 
UV/X-ray luminosities and spectra will 
yield novel insight into the radiative efficiency of the accretion.

\vspace{0.7\baselineskip} 

 {\bf What is the Structure of the Milky Way?}
  Observational studies of the mass and density profile of our Galaxy
  provide key benchmarks for other galaxies, testing
  our knowledge of galaxy assembly.  
  GW observations will offer unique insight from DWD binaries.
  EM studies of these dim
  systems will be mostly confined within
  $\lsim$ kpc from us, even with LSST. Their GW signal, though, remains
  loud and clear from the bulge -- home of most WDs --- and beyond, even to
  the outskirts of the Local Group for the most massive systems \cite{korolk18}.  LISA's numerous resolved detections, and unresolved populations, will
  provide a tomographic picture of the 
  Galaxy~\citep{bena06,adam12,korolr18}. The full picture emerges jointly with optical
  kinematic properties of DWD systems and/or other, more distant
  tracers (e.g. evolved stars), complementing the positional information
  from GWs, to unveil the shape and total dynamical mass of the Galaxy.

\section{\label{sec:Approaches} Approaches to Multimessenger studies}

The most direct multimessenger exchange is to observe a LISA source or
event with EM instruments. Unlike for  LIGO/VIRGO, the
leisurely timescale of a SMBHB binary chirping across the mHz bands
(for weeks or months) enables EM searches for precursors,
flashes or bursts during the merger itself, and afterglows on
timescales from days to years.  Alternatively, we can collect
and use EM information about the same source population (such as SMBH
binaries too wide to be in the LISA band, but producing periodic EM
variability on longer timescales of weeks to years), or even
information about a distinct but astrophysically connected population
(such as the stellar binaries that would necessarily accompany a
population of compact DWDs).
Here we describe some specific examples of joint EM+GW observations
with potential to yield new science, first for massive BH binaries,
and then for compact stellar objects.


{\bf Massive Black Hole Binaries: {}}

{\em Precursor observations.}  LISA can
localize a typical $10^6 \msun$ binary on the sky to within a
few deg$^2$ between a $\sim$week and a few months prior to the merger at $z=1-2$,
or perhaps a $\sim$day at
$z=3$ \citep{KHM2008,LangHughes2008,McWilliams+2011}.  This will
enable a precursor search with a large-FOV
telescope, by monitoring the LISA localization area for days to weeks, covering
up to hundreds of pre-merger orbital cycles~\cite{DalCanton+2019}. 
A promising signal is a quasi-periodic EM ``chirp'', tracking the phase
of the GWs~\citep{Haiman2017,Tang+2017}.  Because of copious shock-heating, gas near the BHs 
is expected to be unusually hot
\citep{Bode+2010,ShiKrolik2012,Roedig+2014,Farris+2015a,Farris+2015b,Bowen+2017,Tang+2018}.
The corresponding UV/X-ray emission would have different (harder)
spectra, with possible signatures of a disk cavity~\citep{Roedig+2014},
and display periodicity
on the orbital
timescale of an $\sim$hour to $\sim$minutes (a month to a day prior to merger)~\cite{Noble+2012}. Doppler effects could 
increase variability over time, tracking the GW
chirp~\citep{Haiman2017,Schnittman+2018}, while self-lensing 
would imprint characteristic, periodically
recurring spikes on the EM light-curve~\cite{Haiman2017,D'Orazio+2018}.
Precursor observations are possible even before LISA is turned on.
A long-lasting SMBH binary can display  
periodicity,
and/or secular changes in its spectrum and overall luminosity, decades
prior to the LISA observations.  This requires pre-existing
time-domain surveys, such as 
LSST in the
optical/IR, of a large fraction of the sky.  Ideally, the
cadence should be sufficient to discover periodicity on the
several-hour orbital timescale, and the depth sufficient to
to detect SMBHs from $z=2-3$.  As argued above, wide UV/X-ray
time-domain surveys, complementing the planned optical/IR surveys, would probe
this crucial pre-merger evolution.

{\em Coincident observations.}  The merger
event itself will produce the most energetic burst of GWs, and may be
accompanied by a luminous X-ray flare from the tidal `squeezing' of
gas~\citep{Chang+2010,Baruteau+2012,TazzariLodato2015,Cerioli+2016,Fontecilla+2017,Pereira+2019},
EM signatures of the turn-on of a relativistic jet, or other flares
from any direct coupling between the GWs and the surrounding
plasma~\cite{Kelly+2017}. Typical sky-localization accuracies for bright SMBH binaries will be a few $\sim$arcmin$^2$.
While the month-long precursor observations will require tiling several deg$^2$ of sky,
and will be feasible only with large-FOV instruments, the flares accompanying the merger
could be targeted by a far wider range of instruments.

{\em After-glows of massive BH mergers.}  After the merger of a SMBH
binary, any circumbinary gas is expected to develop strong prompt
shocks, due to the effectively instant mass-loss and center-of-mass
recoil of the remnant BH.  These should lead to a bright afterglow,
whose nature depends on the amount and configuration of the nearby
gas.  In the case of a thin circumbinary disk, the afterglow should
display a characteristic increase in both spectral hardness and
overall brightness, on time-scales of weeks to
years~\citep{lfh08,sk08,sb08,oneill,megevand,Corrales+2010,rossi2010,megevand2}.
Searching for these after-glows will be possible with a yet larger
range of telescopes; in many cases, the host galaxy of the source will
already have been securely identified.

{\em Observing related populations.}  LISA
binaries represent only the last few years of a merger, i.e.  the
``tip of the iceberg'' of a much longer-lived population of wider
SMBH binaries.  EM observations can map out the demography of
these binaries. Several dozen periodic ($P\sim$ yr) candidate AGN have already been identified in
large optical time-domain  surveys~\cite{Graham2015,Charisi2016}, though a confirmed binary SMBH remains elusive.  
%
%
Future, superior time-domain
surveys (better cadence allowing a search for
shorter periods, homogeneous sampling allowing better separation of
red noise, higher sensitivity allowing a search for lower-mass BHs relevant for LISA) could identify a population of massive BH binaries
approaching, or already in the GW-driven regime~\cite{HKM2009}.
Such surveys can be performed independently
of LISA
in any EM band 
from radio to X-rays.

Many SMBH binaries
may be bright.  A unique quasar may exist in the 3D LISA error
box \cite{Kocsis+06}.  However, even if no unique source is
identified, a Hubble diagram can be done statistically, by each
galaxy in the LISA error box ``voting'' for a source redshift and corresponding
Hubble parameter.  For a clustered host population, the true value of
$H(z)$ will emerge statistically \cite{MacLeodHogan2008,Nair2018}.
Deep galaxy surveys, such as by EUCLID, LSST, WFIRST or LUVOIR, will be important as benchmarks
for identifying host galaxies. This kind of
analysis needs either an all-sky galaxy/quasar catalog complete down to the fluxes of the expected LISA hosts, or else
follow-up surveys will need to be performed for the most promising
LISA sources.
  



{\bf Compact Stellar Objects: {}}

{\em EM observations before LISA.}
Large photometric surveys such as ASAS-SN, ATLAS, CRTS, Gaia, OGLE and PTF/ZTF have shown the potential for identifying short-period variable objects, yielding catalogs of several $\times10^5$ Galactic variables~\cite[e.g.][]{breedt14,jayasinghe+18, heinze+18}. Short-period DWDs show up through brightness variations on the orbital timescale (due to eclipses or tidal deformation of the components). Supplementing these with planned surveys such as BlackGEM, eRosita, LSST and WINTER can yield a multi-wavelength understanding of the EM properties of these events,
which can be matched to future GW detections.
Accreting NSs and the tightest semi-detached WDs are strong X-ray emitters: two of the tightest known binaries, HM Cnc ($P_{\mathrm{orb}}$=5.4min) and V407 Vul ($P_{\mathrm{orb}}$=9.5min), are semi-detached AM CVn systems and are among the loudest LISA verification binaries. Both were selected from the ROSAT survey. eROSITA will provide much deeper observations 
and a complete census of the Galactic population of accretion-powered X-ray binaries down to luminosities of $10^{33-34}$ erg~s$^{-1}$, well in the regime of LISA. We expect that every AM CVn type system within a few kpc and $P_{\rm orb}<10$ min, as well as every ultracompact X-ray binary in the Galaxy will be detected by eROSITA.

{\em EM observations after LISA.}
With LISA providing the approximate sky position and precise orbital period, EM observers will aim to identify the counterpart. The narrowed parameter space will enable deeper searches using orbitally phased co-added images. These follow-ups will provide a wider range of systems to study accretion, resonance locking, tidal heating, etc., than optical surveys alone are likely to find.

\section{Making it Happen}


In various ways, science goals like these rely on \emph{integrating} mHz GW science within the broader astronomy community. It would be careless to passively expect diverse research projects and teams to naturally self-assemble. Rather, opportunities are likely to be missed without a vigorous program of coordinated multidisciplinary investment, beginning long before launch.  Realizing the science potential of LISA will require steady focus on these goals through dedicated multidisciplinary programs.
A broad community-oriented research center for mHz gravitational-wave astronomy is an ideal way to coordinate the necessary investment. Following the CXC and STScI as examples, the center would educate the new user community, and bring theorists, observers, GW instrument builders and data analysis teams together to make the most of this extraordinary science opportunity.

\bibliographystyle{spphys}       

\bibliography{refs}

\end{document}